\documentclass[journal,twocolumn,10pt]{IEEEtran}
\usepackage{amsmath,amssymb}
\usepackage{paralist}
\usepackage{bm}
\usepackage{epsfig, epstopdf}
\usepackage{graphics}
\usepackage{multirow, rotating}
\usepackage{pstricks}
\usepackage{subfigure}
\usepackage{booktabs}

\newtheorem{theorem}{Theorem}
\newtheorem{lemma}{Lemma}
\newtheorem{definition}{Definition}

\newcommand{\A}{\mathcal{A}}
\newcommand{\M}{\mathcal{M}}
\newcommand{\F}{\mathcal{F}}
\renewcommand{\S}{\mathcal{S}}

\newcommand{\Neighb}{\mathcal{N}}
\newcommand{\C}{\mathbb{C}}

\newcommand{\coord}[1]{\mathbf{#1}}

\newcommand{\Span}{\mathbf{span}}

\newcommand{\Nodes}{\mathcal{V}}
\newcommand{\Adj}{\mathbf{A}}
\DeclareMathOperator{\rank}{rank}
\DeclareMathOperator{\Vm}{\mathbf{V}}
\DeclareMathOperator{\Jm}{\mathbf{J}}
\DeclareMathOperator{\ZeroMatrix}{\mathbf{0}}
\DeclareMathOperator{\OneVector}{\mathbf{1}}

\DeclareMathOperator{\FT}{\mathbf{F}}

\DeclareMathOperator{\Circ}{\mathbf{C}}
\DeclareMathOperator{\DFT}{\mathbf{DFT}}
\DeclareMathOperator{\Id}{\mathbf{I}}

\newlength{\whitespacelength}
\newcommand{\ra}[1]{\renewcommand{\arraystretch}{#1}}

\newcommand{\DSPG}{$\mbox{DSP}_{\mbox{\scriptsize G}}$}

\title{Discrete Signal Processing on Graphs}
\author{Aliaksei Sandryhaila$^1$ and Jos\'{e} M.~F. Moura$^1$
}

\author{Aliaksei Sandryhaila,~\IEEEmembership{Member,~IEEE}
    and Jos\'{e} M.~F. Moura,~\IEEEmembership{Fellow,~IEEE}
 \thanks{Copyright (c) 2012 IEEE. Personal use of this material is permitted.
    However, permission to use this material for any other purposes must be
    obtained from the IEEE by sending a request to pubs-permissions@ieee.org.}%
  \thanks{This work was supported in part by AFOSR grant FA95501210087. A.~Sandryhaila and J.~M.~F.~Moura are with the Department of Electrical and Computer
    Engineering, Carnegie Mellon University, Pittsburgh, PA 15213-3890. Ph:~(412)268-6341; fax:~(412)268-3890. Email: asandryh@andrew.cmu.edu, moura@ece.cmu.edu.}
}

\begin{document}

\maketitle

\begin{abstract}
In social settings, individuals interact through webs of relationships. Each individual is a node in a complex network (or graph) of interdependencies and generates data, lots of data. We label the data by its source, or formally stated, we \textit{index} the data by the nodes of the graph. The resulting signals (data indexed by the nodes) are far removed from time or image signals indexed by well ordered time samples or pixels. DSP, discrete signal processing, provides a comprehensive, elegant, and efficient methodology to describe, represent, transform, analyze, process, or synthesize these well ordered time or image signals. This paper extends to \textit{signals on graphs} DSP and its basic tenets, including filters, convolution, $z$-transform, impulse response, spectral representation, Fourier transform, frequency response, and illustrates \textit{DSP on graphs} by classifying blogs, linear predicting and compressing data from irregularly located weather stations, or predicting behavior of customers of a mobile service provider.
\end{abstract}

\textbf{Keywords}: Network science, signal processing, graphical models, Markov random fields, graph Fourier transform.
\section{Introduction}
\label{sec:Intro}
There is an explosion of interest in processing and analyzing large datasets collected in very different settings, including social and economic networks, information networks, internet and the world wide web, immunization and epidemiology networks, molecular and gene regulatory networks, citation and coauthorship studies, friendship networks, as well as physical infrastructure networks like sensor networks, power grids, transportation networks, and other networked critical infrastructures. We briefly overview some of the existing work.

Many authors focus on the underlying \textit{relational structure} of the data by:
 \begin{inparaenum}[1)]
 \item inferring the structure from community relations and friendships, or from perceived alliances between agents as abstracted through game theoretic models \cite{chamley-2004,Jackson:08};
 \item quantifying the connectedness of the world; and
 \item determining the relevance of particular agents, or studying the strength of their interactions.
 \end{inparaenum}
Other authors are interested in the \textit{network function} by quantifying the impact of the network structure on the diffusion of disease, spread of news and information, voting trends, imitation and social influence, crowd behavior, failure propagation, global behaviors developing from seemingly random local interactions~\cite{Jackson:08,easleykleinberg-2010,newman-2010}. Much of these works either develop or assume network models that capture the interdependencies among the data and then analyze the structural properties of these networks.  Models often considered may be deterministic like complete or regular graphs, or random like the Erd\H{o}s-R\'enyi and Poisson graphs, the configuration and expected degree models, small world or scale free networks~\cite{Jackson:08,newman-2010}, to mention a few. These models are used to quantify network characteristics, such as connectedness, existence and size of the giant component, distribution of component sizes, degree and clique distributions, and node or edge specific parameters including clustering coefficients, path length, diameter, betweenness and closeness centralities.

Another body of literature is concerned with inference and learning from such large datasets. Much work falls under the generic label of graphical models~\cite{Whittaker:90,Lauritzen:96,Jensen:01,Jordan:04,Wainwright:08,Koller:09}. In graphical models, data is viewed as a family of random variables indexed by the nodes of a graph, where the graph captures probabilistic dependencies among data elements. The random variables are described by a family of joint probability distributions. For example, directed (acyclic) graphs~\cite{Edwards:00,BangJensen:09} represent Bayesian networks where each random variable is independent of others given the variables defined on its parent nodes. Undirected graphical models, also referred to as Markov random fields~\cite{kindermannsnell-1980,willsky-2002}, describe data where the variables defined on two sets of nodes separated by a boundary set of nodes are statistically independent given the variables on the boundary set. A key tool in graphical models is the Hammersley-Clifford theorem~\cite{kindermannsnell-1980,Besag1974,hammersleyhandscomb}, and the Markov-Gibbs equivalence that, under appropriate positivity conditions, factors the joint distribution of the graphical model as a product of potentials defined on the cliques of the graph. Graphical models exploit this factorization and the structure of the indexing graph to develop efficient algorithms for inference by controlling their computational cost. Inference in graphical models is generally defined as finding from the joint distributions lower order marginal distributions, likelihoods, modes, and other moments of individual variables or their subsets. Common inference algorithms include belief propagation and its generalizations, as well as other message passing algorithms. A recent block-graph algorithm for fast approximate inference, in which the nodes are non-overlapping clusters of nodes from the original graph, is in~\cite{vatsmoura-tsp2012}. Graphical models are employed in many areas; for sample applications, see~\cite{jordansudderthwainrightwillsky-2010} and references therein.

Extensive work is dedicated to discovering efficient data representations for large high-dimensional
data~\cite{Tenenbaum:00,Roweis:00,Belkin:03,Donoho:03}. Many of these works use spectral graph theory and the graph Laplacian~\cite{Chung:96} to derive low-dimensional representations by projecting the data on a low-dimensional subspace generated by a small subset of the Laplacian eigenbasis. The graph Laplacian approximates the Laplace-Beltrami operator on a compact manifold~\cite{Belkin:02,Belkin:03}, in the sense that if the dataset is large and samples uniformly randomly a low-dimensional manifold then the (empirical) graph Laplacian acting on a smooth function on this manifold is a good discrete approximation that converges pointwise and uniformly to the elliptic Laplace-Beltrami operator applied to this function as the number of points goes to infinity~\cite{heinaudibertluxburg-2005,ginekoltchinskii-2006,heinaudibertluxburg-2007}. One can go beyond the choice of the graph Laplacian by choosing discrete approximations to other continuous operators and obtaining possibly more desirable spectral bases for the characterization of the geometry of the manifold underlying the data. For example, if the data represents a non-uniform sampling of a continuous manifold, a conjugate to an elliptic Schr{\"o}dinger-type operator can be used~\cite{Coifman:05a,Coifman:05b,Coifman:06}.

More in line with our paper, several works have proposed multiple transforms for data indexed by graphs. Examples include regression algorithms~\cite{guestrinbodikthibauxpaskinmadden-ipsn2004}, wavelet decompositions~\cite{ganesangreensteinestrinheidemanngovindan-2005,wagnerbaraniuketal-SSPWorkshop2005,wagnerbaraniuketal-IPSN06, Coifman:06,Hammond:11},
filter banks on graphs~\cite{Narang:10,Narang:12}, de-noising~\cite{wagnerdelouillebaraniuk-2006}, and compression~\cite{Zhu:12}. Some of these transforms focus on distributed processing of data from sensor fields while addressing sampling irregularities due to random sensor placement. Others consider localized processing of signals on graphs in multiresolution fashion by representing data using wavelet-like bases with varying ``smoothness'' or defining transforms based on node neighborhoods. In the latter case, the graph Laplacian and its eigenbasis are sometimes used to define a spectrum and a Fourier transform of a signal on a graph. This definition of a Fourier transform was also proposed for use in uncertainty analysis on graphs~\cite{Agaskar:12a,Agaskar:12}. This graph Fourier transform is derived from the graph Laplacian and restricted to undirected graphs with real, non-negative edge weights, not extending to data indexed by directed graphs or graphs with negative or complex weights.

The algebraic signal processing (ASP) theory~\cite{Pueschel:03a,Pueschel:05e,Pueschel:08a,Pueschel:08b} is a formal, algebraic approach to analyze data indexed by special types of line graphs and lattices. The theory uses an algebraic representation of signals and filters as polynomials to derive fundamental signal processing concepts. This framework has been used for discovery of fast computational algorithms for discrete signal transforms~\cite{Pueschel:03a, Pueschel:08c, Sandryhaila:11a}. It was extended to multidimensional signals and nearest neighbor graphs~\cite{Pueschel:07,Sandryhaila:12} and applied in signal compression~\cite{Sandryhaila:12a, Sandryhaila:12b}. The framework proposed in this paper generalizes and extends the ASP to signals on arbitrary graphs.


\subsection*{Contribution}
%
Our goal is to develop a linear discrete signal processing (DSP) framework and corresponding tools for datasets arising from social, biological, and physical networks. DSP has been very successful in processing time signals (such as speech, communications, radar, or econometric time series), space-dependent signals (images and other multidimensional signals like seismic and hyperspectral data), and time-space signals (video). We refer to data \textit{indexed} by nodes of a graph as a \textit{graph signal} or simply signal and to our approach as \emph{DSP on graphs} (\DSPG)\footnote{The term ``signal processing for graphs'' has been used in~\cite{Miller:10, Miller:11} in reference to graph structure analysis and subgraph detection. It should not be confused with our proposed DSP framework, which aims at the analysis and processing of \emph{data} indexed by the nodes of a graph.}. We introduce the basics of linear\footnote{We are concerned with linear operations; in the sequel, we refer only to \DSPG\ but have in mind that we are restricted to linear \DSPG.} \DSPG, including the notion of a shift on a graph, filter structure, filtering and convolution, signal and filter spaces and their algebraic structure, the graph Fourier transform, frequency, spectrum, spectral decomposition, and impulse and frequency responses. With respect to other works, ours is a deterministic framework to signal processing on graphs rather than a statistical approach like graphical models. Our work is an extension and generalization of the traditional DSP, and generalizes the ASP theory~\cite{Pueschel:03a,Pueschel:05e,Pueschel:08a,Pueschel:08b} and its extensions and applications~\cite{Sandryhaila:12,Sandryhaila:12a, Sandryhaila:12b}. We emphasize the contrast between the \DSPG\ and the approach to the graph Fourier transform that takes the graph Laplacian as a point of departure~\cite{ganesangreensteinestrinheidemanngovindan-2005,wagnerdelouillebaraniuk-2006,Narang:10,Hammond:11,Zhu:12,Agaskar:12}.
In the latter case, the Fourier transform on graphs is given by the eigenbasis of the graph Laplacian. However, this definition is not applicable to directed graphs, which often arise in real-world problems, as demonstrated by examples in Section~\ref{sec:Applications}, and graphs with negative weights. In general, the graph Laplacian is a second-order operator for signals on a graph, whereas an adjacency matrix is a first-order operator. Deriving a graph Fourier transform from the graph Laplacian is analogous in traditional DSP to restricting signals to be even (like correlation sequences) and Fourier transforms to represent power spectral densities of signals.
Instead, we demonstrate that the graph Fourier transform is properly defined through the Jordan normal form and generalized eigenbasis of the adjacency matrix\footnote{
Parts of this material also appeared in~\cite{Sandryhaila:13a,Sandryhaila:13b}. In this paper, we present a complete theory with all derivations and proofs.}.
Finally, we illustrate the \DSPG\ with applications like classification, compression, and linear prediction for datasets that include blogs, customers of a mobile operator, or collected by a network of irregularly placed weather stations.
%

\section{Signals on Graphs}
\label{sec:Signals_on_graphs}
Consider a dataset with $N$ elements, for which some \emph{relational} information about its data elements is known. Examples include preferences of individuals in a social network and their friendship connections, the number of papers published by authors and their coauthorship relations, or topics of online documents in the World Wide Web and their hyperlink references.
This information can be represented by a graph $G=(\Nodes,\Adj)$, where $\Nodes=\{v_0,\ldots,v_{N-1}\}$ is the set of nodes and $\Adj$ is the weighted\footnote{Some literature defines the adjacency matrix $\Adj$ of a graph $G=(\Nodes,\Adj)$ so that $\Adj_{n,m}$ only takes values of 0 or 1, depending on whether there is an edge from $v_m$ to $v_n$, and specifies edge weights as a function on pairs of nodes. In this paper, we incorporate edge weights into~$\Adj$.} adjacency matrix of the graph. Each dataset element corresponds to node $v_n$, and each weight $\Adj_{n,m}$ of a \emph{directed} edge from $v_m$ to $v_n$ reflects the degree of relation of the $n$th element to the $m$th one. Since data elements can be related to each other differently, in general, $G$ is a \emph{directed, weighted} graph. Its edge weights $\Adj_{n,m}$ are not restricted to being nonnegative reals; they can take arbitrary real or complex values (for example, if data elements are negatively correlated). The set of indices of nodes connected to $v_n$ is called the \emph{neighborhood} of $v_n$ and denoted by $\Neighb_n=\{m \mid \Adj_{n,m}\neq 0 \}$.

Assuming, without a loss of generality, that dataset elements are complex scalars, we define a \emph{graph signal} as a map from the set $\Nodes$ of nodes into the set of complex numbers $\C$:
\vspace{-2mm}
\begin{eqnarray}
\nonumber
\coord{s} & : & \Nodes\,\rightarrow\,\C, \\
\label{eq:s}
&& v_n\,\mapsto\,s_n.
\vspace{-2mm}
\end{eqnarray}
Notice that each signal is isomorphic to a complex-valued vector with $N$ elements. Hence, for simplicity of discussion, we write graph signals as vectors
$\coord{s} = \begin{pmatrix} s_0 & s_1 & \ldots & s_{N-1} \end{pmatrix}^T$, but remember that each element $s_n$ is \emph{indexed} by node $v_n$ of a given representation graph $G=(\Nodes,\Adj)$, as defined by~\eqref{eq:s}.
The space $\S$ of graphs signals~\eqref{eq:s} then is identical to $\C^N$.

\begin{figure}
  \begin{center}
    \subfigure[Time series]{\label{fig:graphs_time}\includegraphics[scale=0.4]{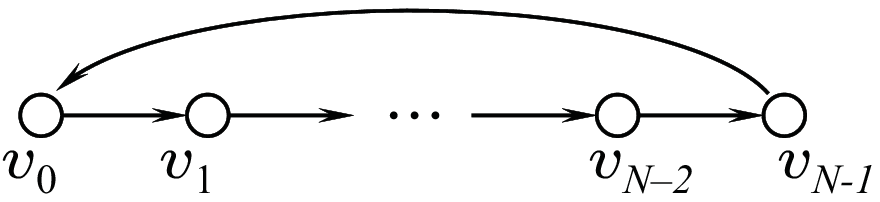}}
    \hspace{5mm}
    \subfigure[Digital image]{\label{fig:graphs_image}\includegraphics[scale=0.3]{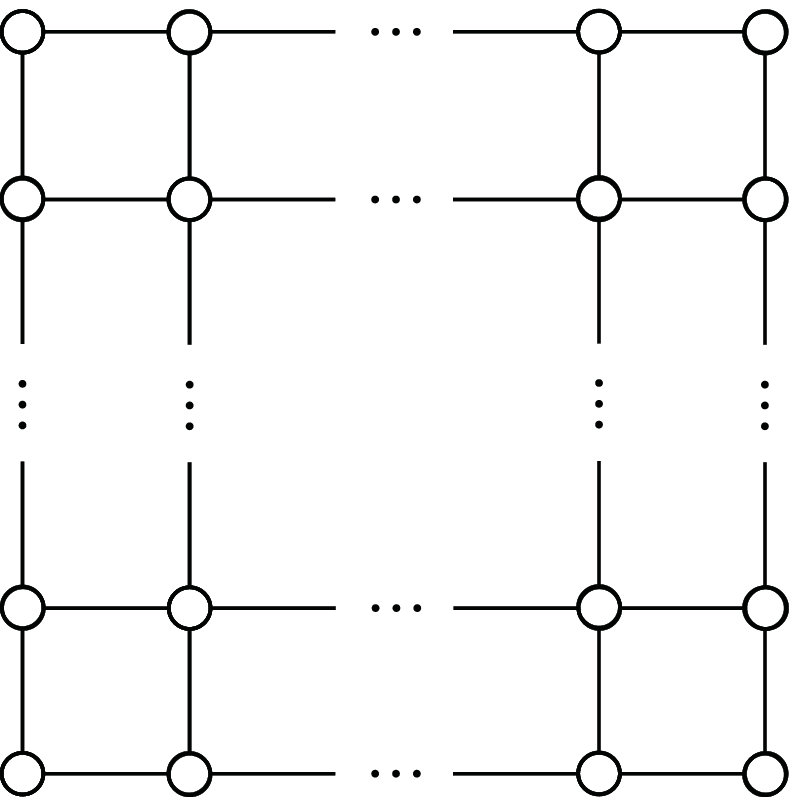}} \\
    \subfigure[Sensor field]{\label{fig:graphs_weather}\includegraphics[scale=0.18]{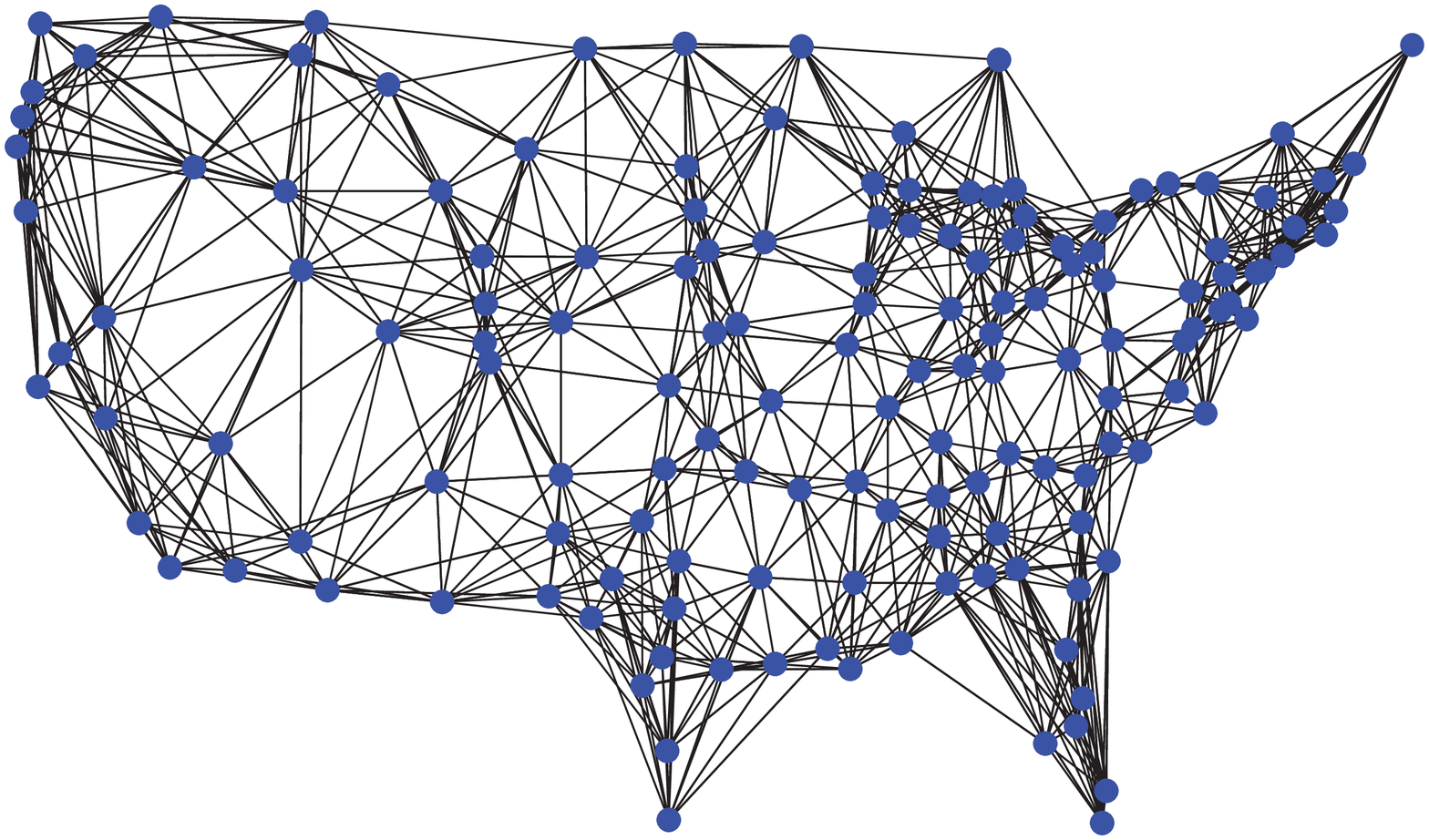}}
    \subfigure[Hyperlinked documents]{\label{fig:graphs_blogs}\includegraphics[scale=0.18]{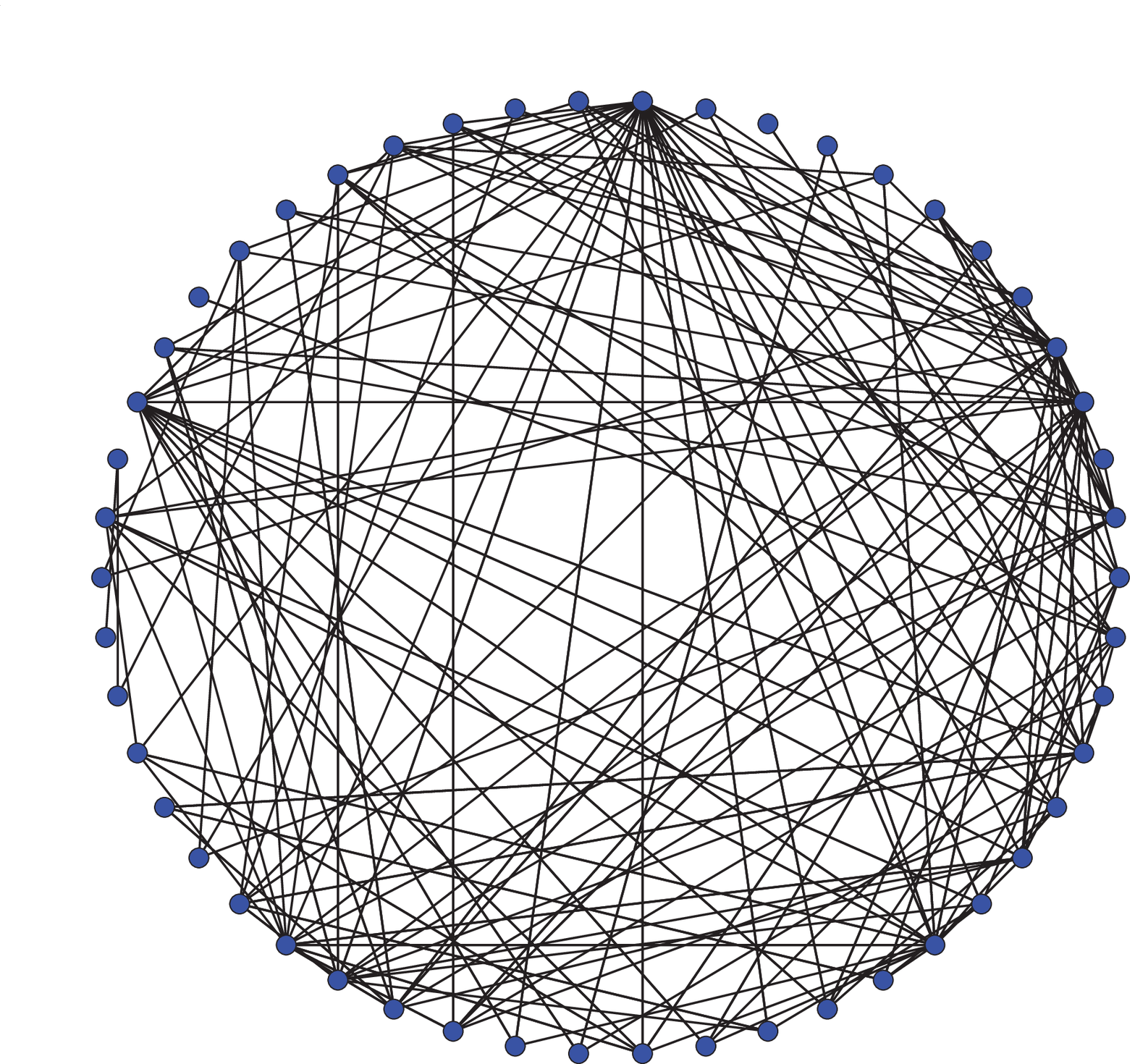}}
  \end{center}
  \vspace{-2mm}
\caption{\label{fig:graphs} Graph representations for different datasets (graph signals.)}
\vspace{-5mm}
\end{figure}
We illustrate representation graphs with examples shown in Fig.~\ref{fig:graphs}. The directed cyclic graph in Fig.~\ref{fig:graphs_time} represents a finite, periodic discrete time series~\cite{Pueschel:08a}. All edges are directed and have the same weight~$1$, reflecting the causality of a time series; and the edge from $v_{N-1}$ to $v_0$ reflects its periodicity. The two-dimensional rectangular lattice in Fig.~\ref{fig:graphs_image} represents a general digital image. Each node corresponds to a pixel, and each pixel value (intensity) is related to the values of the four adjacent pixels. This relation is symmetric, hence all edges are undirected and have the same weight, with possible exceptions of boundary nodes that may have directed edges and/or different edge weights, depending on boundary conditions~\cite{Pueschel:08b}. Other lattice models can be used for images as well~\cite{Pueschel:07}. The graph in Fig.~\ref{fig:graphs_weather} represents temperature measurements from~$150$ weather stations (sensors) across the United States. We represent the relations of temperature measurements by geodesic distances between sensors, so each node is connected to its closest neighbors. The graph in Fig.~\ref{fig:graphs_blogs} represents a set of~$50$ political blogs in the World Wide Web connected by hyperlink references. By their nature, the edges are directed and have the same weights. We discuss the two latter examples in Section~\ref{sec:Applications}, where we also consider a network of customers of a mobile service provider. Clearly, representation graphs depend on prior knowledge and assumptions about datasets. For example, the graph in Fig.~\ref{fig:graphs_blogs} is obtained by following the hyperlinks networking the blogs, while the graph in Fig.~\ref{fig:graphs_weather} is constructed from known locations of sensors under assumption that temperature measurements at nearby sensors have highly correlated temperatures.

\section{Filters on Graphs}
\label{sec:Filters_on_Graphs}
%
In classical DSP, signals are processed by \emph{filters}---systems that take a signal as input and produce another signal as output. We now develop the equivalent concept of \emph{graph filters} for graph signals in \DSPG. We consider only linear, shift-invariant filters, which are a generalization of linear time-invariant filters used in DSP for time series. This section uses Jordan normal form and characteristic and minimal polynomials of matrices; these concepts are reviewed in Appendix~A. The use of Jordan decomposition is required since for many real-world datasets the adjacency matrix~$\Adj$ is not diagonalizable. One example is the blog dataset, considered in Section~\ref{sec:Applications}.
%
\subsection*{Graph Shift}
\label{subsec:shift}
%
In classical DSP, the basic building block of filters is a special filter $x=z^{-1}$ called the~\emph{time shift} or \emph{delay}~\cite{Oppenheim:99}. This is the simplest non-trivial filter that delays the input signal~$\coord{s}$ by one sample, so that the $n$th sample of the output is $\widetilde{s}_n=s_{n-1\mod N}$.
Using the graph representation of finite, periodic time series in Fig.~\ref{fig:graphs_time}, for which the adjacency matrix is the $N\times N$ circulant matrix $\Adj=\Circ_N$,
with weights~\cite{Pueschel:05e,Pueschel:08a}
\begin{equation}
\label{eq:time_shift_matrix}
\Adj_{n,m} =
\begin{cases}
1,& \text{   if   } n-m = 1 \mod N  \\
0,& \text{   otherwise  }
\end{cases},
\end{equation}
we can write the time shift operation as
\begin{equation}
\label{eq:shift}
\coord{\tilde{s}} = \Circ_N\coord{s} = \Adj\coord{s}.
\end{equation}

In \DSPG, we extend the notion of the shift~\eqref{eq:shift} to general graph signals~$\coord{s}$ where the relational dependencies among the data are represented by an arbitrary graph $G=(\Nodes,\Adj)$. We call the operation~\eqref{eq:shift} the \emph{graph shift}. It is realized by replacing the sample~$s_n$ at node~$v_n$ with the weighted linear combination of the signal samples at its neighbors:
\begin{equation}
\label{eqn:shiftsamplegenericgraph}
\tilde{s}_n = \sum_{m=0}^{N-1} \Adj_{n,m} s_m = \sum_{m\in\Neighb_n} \Adj_{n,m} s_m.
\end{equation}
Note that, in classical DSP, shifting a finite signal requires one to consider boundary conditions. In \DSPG, this problem is implicitly resolved, since the graph~$G=(\Nodes,\Adj)$ explicitly captured the boundary conditions.

%
\subsection*{Graph Filters}
\label{subsec:filters}
%
Similarly to traditional DSP, we can represent filtering on a graph using matrix-vector multiplication. Any system $\coord{H}\in\C^{N\times N}$, or \emph{graph filter}, that for input $\coord{s}\in\S$ produces output $\coord{H}\coord{s}$ represents a \emph{linear} system, since the filter's output for a linear combination of input signals equals the linear combination of outputs to each signal:
$$
\coord{H}(\alpha\coord{s}_1+\beta\coord{s}_2) = \alpha\coord{H}\coord{s}_1+\beta\coord{H}\coord{s}_2.
$$
Furthermore, we focus on \emph{shift-invariant} graph filters, for which applying the graph shift to the output is equivalent to applying the graph shift to the input prior to filtering:
\begin{equation}
\label{eq:shift-invariance}
\Adj\big( \coord{H}\coord{s} \big) = \coord{H} \big( \Adj\coord{s} \big).
\end{equation}

The next theorem establishes that all linear, shift-invariant graph filters are given by \emph{polynomials} in the shift~$\Adj$.
\begin{theorem}
\label{thm:shift_invariant_polynomials}
Let~$\Adj$ be the graph adjacency matrix and assume that its characteristic and minimal polynomials are equal: $p_\Adj(x) = m_\Adj(x)$. Then, a graph filter~$\coord{H}$ is linear and shift invariant if and only if~(iff) $\coord{H}$ is a \emph{polynomial} in the graph shift $\Adj$, i.e., iff there exists a polynomial
\begin{equation}
\label{eq:pol_hx}
h(x) = h_0 + h_1 x + \ldots + h_L x^L
\end{equation}
with possibly complex coefficients $h_\ell\in\C$, such that:
\begin{equation}
\label{eq:filter}
\coord{H} = h(\Adj) = h_0\Id + h_1\Adj + \ldots + h_L\Adj^L.
\end{equation}
\end{theorem}
\begin{IEEEproof}
Since the shift-invariance condition~\eqref{eq:shift-invariance} holds for all graph signals $\coord{s}\in\S=\C^N$, the matrices $\Adj$ and $\coord{H}$ commute: $\Adj \coord{H} = \coord{H} \Adj$. As $p_\Adj(x) = m_\Adj(x)$, all eigenvalues of $\Adj$ have exactly one eigenvector associated with them, \cite{Gantmacher:59,Lancaster:85}. Then, the graph matrix $\coord{H}$ commutes with the shift~$\Adj$ iff it is a polynomial in~$\Adj$ (see Proposition 12.4.1 in~\cite{Lancaster:85}).
\end{IEEEproof}

Analogous to the classical DSP, we call the coefficients $h_\ell$ of the polynomial $h(x)$ in~\eqref{eq:pol_hx} the graph filter \emph{taps}.

\subsection*{Properties of Graph Filters}

Theorem~\ref{thm:shift_invariant_polynomials} requires the equality of the characteristic and minimal polynomials~$p_\Adj(x)$ and~$m_\Adj(x)$. This condition does not always hold, but can be successfully addressed through the concept of \textit{equivalent} graph filters, as defined next.
\begin{definition}
\label{def:equivalent_filters}
Given any shift matrices $\Adj$ and $\widetilde{\Adj}$, filters $h(\Adj)$ and $g(\widetilde{\Adj})$ are called \emph{equivalent} if for all input signals $\coord{s}\in\S$ they produce equal outputs: $h(\Adj)\coord{s} = g(\widetilde{\Adj})\coord{s}.$
\end{definition}

Note that, when no restrictions are placed on the signals, so that $\S=\C^N$, Definition~\ref{def:equivalent_filters} is equivalent to requiring $h(\Adj) = g(\widetilde{\Adj})$ as matrices. However, if additional restrictions exist, filters may not necessarily be equal as matrices and still produce the same output for the considered set of signals.

It follows that, given an arbitrary $G=(\Nodes,\Adj)$ with $p_\Adj(x) \neq  m_\Adj(x)$, we can consider another graph $\widetilde{G}=(\Nodes,\widetilde{\Adj})$ with the same set of nodes $\Nodes$ but potentially different edges and edge weights, for which $p_{\widetilde{\Adj}}(x) = m_{\widetilde{\Adj}}(x)$ holds true. Then graph filters on $G$ can be expressed as equivalent filters on $\widetilde{G}$, as described by the following theorem (proven in Appendix B).
\begin{theorem}
\label{thm:shift_via_shift}
For any matrix $\Adj$ there exists a matrix $\widetilde{\Adj}$ and polynomial $r(x)$, such that $\Adj = r(\widetilde{\Adj})$ and $p_{\widetilde{\Adj}}(x) = m_{\widetilde{\Adj}}(x).$
\end{theorem}

As a consequence of Theorem~\ref{thm:shift_via_shift}, any filter on the graph $G=(\Nodes,\Adj)$ is equivalent to a filter on the graph $\widetilde{G}=(\Nodes,\widetilde{\Adj})$, since $h(\Adj)=h(r(\widetilde{\Adj})) = (h\circ r)(\widetilde{\Adj})$, where $h\circ r$ is the composition of polynomials $h$ and $r$ and thus is a polynomial.
Thus, the condition $p_\Adj(x)=m_\Adj(x)$ in Theorem~\ref{thm:shift_invariant_polynomials} can be assumed to hold for any graph $G=(\Nodes,\Adj)$. Otherwise, by Theorem~\ref{thm:shift_via_shift}, we can replace the graph by another $\widetilde{G}=(\Nodes,\widetilde{\Adj})$ for which the condition holds and assign~$\widetilde{\Adj}$ to~$\Adj$.

The next result demonstrates that we can limit the number of taps in any graph filter.
\begin{theorem}
\label{thm:equivalence}
Any graph filter~\eqref{eq:filter} has a unique equivalent filter on the same graph with at most $\deg m_\Adj(x) = N_\Adj$ taps.
\end{theorem}
\begin{IEEEproof}
Consider the polynomials $h(x)$ in~\eqref{eq:pol_hx}. By polynomial division, there exist unique polynomials $q(x)$ and $r(x)$:
\vspace{-2mm}
\begin{equation}
\label{eq:equivalence}
h(x) = q(x)m_\Adj(x) + r(x),
\vspace{-1mm}
\end{equation}
where $\deg r(x) < N_\Adj$. Hence, we can express~\eqref{eq:filter} as
$$
h(\Adj) = q(\Adj)m_\Adj(\Adj) + r(\Adj) = q(\Adj)\ZeroMatrix_N + r(\Adj) = r(\Adj).
$$
Thus, $h(\Adj) = r(\Adj)$ and $\deg r(x) < \deg m_\Adj(x)$. 
\end{IEEEproof}

As follows from Theorem~\ref{thm:equivalence}, all linear, shift-invariant filters~\eqref{eq:filter} on a graph $G=(\Nodes,\Adj)$ form a vector space
\begin{align}
\label{eq:filter_space}
\F = \left\{ \coord{H}:\:\:\left.\coord{H}{}=\sum_{\ell=0}^{N_\Adj-1} h_\ell \Adj^\ell \right| h_\ell \in \C \right\}.
\end{align}
Moreover, addition and multiplication of filters in $\F$ produce new filters that are equivalent to filters in $\F$. Thus, $\F$ is closed under these operations, and has the structure of an algebra~\cite{Pueschel:05e}. We discuss it in detail in Section~\ref{sec:Algebraic_model}.

Another consequence of Theorem~\ref{thm:equivalence} is that the inverse of a filter on a graph, if it exists, is also a filter on the same graph, i.e., it is a polynomial in~\eqref{eq:filter_space}.
\begin{theorem}
\label{thm:inverse_filter}
A graph filter $\coord{H}=h(\Adj)\in\F$ is invertible iff polynomial $h(x)$
satisfies $h(\lambda_m)\neq 0$ for all distinct eigenvalues $\lambda_0,\ldots,\lambda_{M-1}$, of $\Adj$.
Then, there is a unique polynomial $g(x)$ of degree $\deg g(x) < N_\Adj$ that satisfies
\vspace{-1mm}
\begin{equation}
\label{eq:g_inverse_h}
h(\Adj)^{-1}=g(\Adj)\in\F.
\vspace{-1mm}
\end{equation}
\end{theorem}
Appendix~C contains the proof and the procedure for the construction of~$g(x)$.
Theorem~\ref{thm:inverse_filter}  implies that instead of inverting the $N\times N$
matrix $h(\Adj)$ directly we only need to construct a polynomial~$g(x)$ specified by at most $N_{\Adj}$ taps.

Finally, it follows from Theorem~\ref{thm:equivalence} and~\eqref{eq:filter_space} that any graph filter $h(\Adj)\in\F$ is completely specified by its taps $h_0,\cdots,h_{N_\Adj-1}$. As we prove next, in \DSPG, as in traditional DSP, filter taps uniquely determine the \emph{impulse response} of the filter, i.e., its output $\coord{u} = \left( g_0,\ldots,g_{N-1}\right)^T$ for unit impulse input $\bm{\delta}=\left(1,0,\ldots,0\right)^T$, and vice versa.
\begin{theorem}
\label{thm:impulseresponse-1}
The filter taps $h_0,\ldots,h_{N_\Adj-1}$ of the filter~$h(\Adj)$ uniquely determine its impulse response $\coord{u}$.
Conversely, the impulse response $\coord{u}$ uniquely determines the filter taps, provided $\rank\widehat{\Adj} = N_\Adj$, where
$\widehat{\Adj} = \left( \Adj^0\bm{\delta} , \ldots , \Adj^{N_\Adj-1}\bm{\delta} \right).$
\end{theorem}
\begin{IEEEproof}
The first part follows from the definition of filtering: $\coord{u}=h(\Adj)\bm{\delta}=\widehat{\Adj}\coord{h}$ yields the first column of $h(\Adj)$, which is uniquely determined by the taps $\coord{h}=\left(h_0,\ldots,h_{N_\Adj-1}\right)^T$.
Since we assume $p_\Adj(x)=m_\Adj(x),$ then $N=N_\Adj$, and the second part holds if $\widehat{\Adj}$ is invertible, i.e., $\rank\widehat{\Adj} = N_\Adj$.
\end{IEEEproof}

Notice that a relabeling of the nodes $v_0,\ldots,v_{N-1}$ does not change the impulse response. If $\coord{P}$ is the corresponding permutation matrix, then the unit
impulse is $\coord{P}\bm{\delta}$, the adjacency matrix is $\coord{P}\Adj\coord{P}^T$, and the filter becomes $h(\coord{P}\Adj\coord{P}^T) = \coord{P}h(\Adj)\coord{P}^T$.
Hence, the impulse response is simply reordered according to same permutation: $\coord{P}h(\Adj)\coord{P}^T \coord{P}\bm{\delta} = \coord{P}\coord{u}$.

%

\section{Algebraic Model}
\label{sec:Algebraic_model}

So far, we presented signals and filters on graphs as vectors and matrices. An alternative representation exists for filters and signals as polynomials. We call this representation the graph $z$-transform, since, as we show, it generalizes the traditional $z$-transform for discrete time signals that maps signals and filters to polynomials or series in $z^{-1}$. The graph $z$-transform is defined separately for graph filters and signals.


Consider a graph $G =(\Nodes, \Adj)$, for which the characteristic and minimal polynomials of the adjacency matrix coincide: $p_\Adj(x) = m_\Adj(x)$.
The mapping $\Adj\mapsto x$ of the adjacency matrix~$\Adj$ to the indeterminate~$x$ maps the graph filters $\coord{H}=h(\Adj)$ in~$\F$ to polynomials~$h(x)$. By Theorem~\ref{thm:equivalence}, the filter space~$\F$ in~\eqref{eq:filter_space} becomes a \emph{polynomial algebra}~\cite{Pueschel:05e}
\begin{align}
\label{eqn:filter-polyalgebraA}
\A=\C[x]/m_\Adj(x).
\end{align}
This is a space of polynomials of degree less than $\deg m_\Adj(x)$ with complex coefficients that is closed under addition and multiplication of polynomials modulo $m_\Adj(x)$. The mapping $\F\rightarrow \A$, $h(\Adj)\mapsto h(x)$, is a isomorphism of $\C$-algebras~\cite{Pueschel:05e}, which we denote as $\F\cong\A$. We call it the \emph{graph $z$-transform of filters} on graph $G =(\Nodes, \Adj)$.

The signal space~$\S$ is a vector space that is also closed under filtering, i.e., under multiplication by graph filters from $\F$: for any signal $\coord{s}\in\S$ and filter $h(\Adj)$, the output is a signal in the same space: $h(\Adj)\coord{s}\in\S$. Thus, $\S$ is an $\F$-module~\cite{Pueschel:05e}.
As we show next, the \emph{graph $z$-transform of signals} is defined as an isomorphism~\eqref{eq:z_transform_graphs} from~$\S$ to an $\A$-module.
\begin{theorem}
\label{thm:z-transform}
Under the above conditions, the signal space $\S$ is isomorphic to an $\A$-module
\begin{align}
\label{eq:signalmoduleongraphs}
\M {}&= \C[x]/p_\Adj(x) = \left\{s(x)=\sum_{n=0}^{N-1} s_n b_n(x) \right\}
\end{align}
under the mapping
\begin{equation}
\label{eq:z_transform_graphs}
\coord{s} = \left(s_0,\ldots,s_{N-1}\right)^T
\mapsto s(x)=\sum_{n=0}^{N-1} s_n b_n(x).
\end{equation}
The polynomials $b_0(x),\ldots,b_{N-1}(x)$ are linearly independent polynomials of degree at most $N-1$.
If we write
\begin{equation}
\label{eq:vector_b}
\coord{b}(x) = \left( b_0(x),\ldots,b_{N-1}(x)\right)^T,
\end{equation}
then the polynomials satisfy
\begin{equation}
\label{eq:condition_on_basis}
\coord{b}^{(r)}(\lambda_m)=
\begin{pmatrix}
b_0^{(r)}(\lambda_m)&\ldots&b_{N-1}^{(r)}(\lambda_m)
\end{pmatrix}^T
=r!\coord{\tilde{v}}_{m,0,r}
\end{equation}
for $0\leq r<R_{m,0}$ and $0\leq m<M$, where $\lambda_m$ and $\coord{\tilde{v}}_{m,0,r}$ are generalized eigenvectors of $\Adj^T,$
and $b_n^{(r)}(\lambda_m)$ denotes the $r$th derivative of $b_n(x)$ evaluated at $x=\lambda_m$.
Filtering in $\M$ is performed as multiplication modulo $p_\Adj(x)$: if $\coord{\tilde{s}}=h(\Adj)\coord{s}$, then
\begin{equation}
\label{eq:filtering_as_multiplication}
\coord{\tilde{s}} \,\,\mapsto\,\,
\tilde{s}(x)=\sum_{n=0}^{N-1} \tilde{s}_n b_n(x) = h(x)s(x)\mod p_\Adj(x).
\end{equation}
\end{theorem}
\begin{IEEEproof}
Due to the linearity and shift-invariance of graph filters, we only need to prove~\eqref{eq:filtering_as_multiplication} for $h(\Adj)=\Adj.$
Let us write $s(x) = \coord{b}(x)^T\coord{s}$ and $\tilde{s}(x) = \coord{b}(x)^T\coord{\tilde{s}}$, where $\coord{b}(x)$ is given by~\eqref{eq:vector_b}.
Since~\eqref{eq:filtering_as_multiplication} must hold for all $\coord{s}\in\S$, for $h(\Adj)=\Adj$ it is equivalent to
\begin{eqnarray}
\nonumber
&& \coord{b}(x)^T\coord{\tilde{s}} = \coord{b}(x)^T\left(\Adj\coord{s}\right) = \coord{b}(x)^T\left(x\coord{s}\right) \mod p_\Adj(x) \\
\label{eq:condition_on_b}
& \Leftrightarrow & \left(\Adj^T-x\Id\right)\coord{b}(x) = \coord{c} p_\Adj(x),
\end{eqnarray}
where $\coord{c}\in\C^N$ is a vector of constants, since $\deg p_\Adj(x) = N$ and $\deg \left(xb_n(x)\right) \leq N$ for $0\leq n<N$.

It follows from the factorization~\eqref{eq:characteristic_polynomial} of $p_\Adj(x)$ that, for each eigenvalue $\lambda_m$ and $0\leq k<A_m$, the characteristic polynomial satisfies $p_\Adj^{(k)}(\lambda_m)=0$.
By taking derivatives of both sides of~\eqref{eq:condition_on_b} and evaluating at $x=\lambda_m$, $0\leq m < M$, we construct $A_0+\ldots+A_{M-1}=N$ linear equations
\begin{eqnarray*}
\left(\Adj^T-\lambda_m\Id\right)\coord{b}(\lambda_m) &=& 0 \\
\left(\Adj^T-\lambda_m\Id\right)\coord{b}^{(r)}(\lambda_m) &=& r\coord{b}(\lambda_m), \,\, 1\leq r< A_m
\end{eqnarray*}
Comparing these equations with~\eqref{eq:generalized_eigenvector_condition}, we obtain~\eqref{eq:condition_on_basis}.
Since $N$ polynomials $b_n(x)=b_{n,0}+\ldots+b_{n,N-1}x^{N-1}$ are characterized by $N^2$ coefficients $b_{n,k}$, $0\leq n,k<N$,~\eqref{eq:condition_on_basis} is a system of $N$ linear equations with $N^2$ unknowns that can always be solved using inverse polynomial interpolation~\cite{Lancaster:85}.
\end{IEEEproof}

Theorem~\ref{thm:z-transform} extends to the general case $p_\Adj(x) \neq m_\Adj(x)$. By Theorem~\ref{thm:shift_via_shift}, there exists a graph $\widetilde{G} = (\Nodes, \widetilde{\Adj})$ with $p_{\widetilde{\Adj}}(x) = m_{\widetilde{\Adj}}(x)$, such that $\Adj = r(\widetilde{\Adj})$. By mapping $\widetilde{\Adj}$ to $x$, the filter space~\eqref{eq:filter_space} has the structure of the polynomial algebra $\A=\C[x]/m_\Adj(r(x)) = \C[x]/(m_\Adj\circ r)(x))$ and the signal space has the structure of the $\A$-module $\M=\C[x]/p_{\widetilde{\Adj}}(x)$. Multiplication of filters and signals is performed modulo $p_{\widetilde{\Adj}}(x)$. The basis of $\M$ satisfies~\eqref{eq:condition_on_basis}, where $\lambda_m$ and $\coord{v}_{m,d,r}$ are eigenvalues and generalized eigenvectors of $\widetilde{\Adj}.$

\section{Fourier Transform on Graphs}
\label{sec:FT}

After establishing the structure of filter and signal spaces in \DSPG, we define other fundamental DSP concepts, including spectral decomposition, signal spectrum, Fourier transform, and frequency response. They are related to the Jordan normal form of the adjacency matrix $\Adj$, reviewed in Appendix~A.

\subsection*{Spectral Decomposition}

In DSP, spectral decomposition refers to the identification of subspaces $\S_0,\ldots,\S_{K-1}$ of the signal space $\S$ that are \emph{invariant} to filtering, so that, for any signal $\coord{s}_k\in\S_k$ and filter $h(\Adj)\in\F$, the output $\widetilde{\coord{s}}_k= h(\Adj)\coord{s}_k$ lies in the same subspace $\S_k$. A signal $\coord{s}\in\S$ can then be represented as
\begin{equation}
\label{eq:spectral_decomposition_s}
\coord{s} = \coord{s}_0 + \coord{s}_1 + \ldots + \coord{s}_{K-1},
\end{equation}
with each component $\coord{s}_k\in\S_k$. Decomposition~\eqref{eq:spectral_decomposition_s} is uniquely determined for every signal $\coord{s}\in\S$ if and only if:
\begin{inparaenum}[1)]
    \item
    invariant subspaces $\S_k$ have zero intersection, i.e., $\S_k \cap \S_m = \{0\}$ for $k\neq m$;
    \item
    $\dim\S_0 + \ldots + \dim \S_{K-1} = \dim\S = N$; and
    \item
    each $\S_k$ is \emph{irreducible}, i.e., it cannot be decomposed into smaller invariant subspaces.
\end{inparaenum}
In this case, $\S$ is written as a direct sum of vector subspaces
\begin{equation}
\label{eq:direct_sum}
\S = \S_0 \oplus \S_1 \oplus \ldots \oplus \S_{K-1}.
\end{equation}

As mentioned, since the graph may have arbitrary structure, the adjacency matrix~$\Adj$ may not be diagonalizable; in fact, $\Adj$ for the blog dataset (see Section~\ref{sec:Applications}) is not diagonalizable. Hence, we consider the Jordan decomposition~\eqref{eq:Jordan_decomposition} $\Adj=\Vm \Jm \Vm^{-1}$, which is reviewed in Appendix A. Here, $\Jm$ is the Jordan normal form~\eqref{eq:Jordan_normal_form}, and $\Vm$ is the matrix of generalized eigenvectors~\eqref{eq:generalized_eigenvectors_matrix}. Let $\S_{m,d} = \Span\{\coord{v}_{m,d,0},\ldots,\coord{v}_{m,d,R_{m,d}-1} \}$ be a vector subspace of $\S$ spanned by the $d$th Jordan chain of $\lambda_m$. Any signal $\coord{s}_{m,d}\in\S_{m,d}$ has a unique expansion
\begin{eqnarray*}
\coord{s}_{m,d} &=& \hat{s}_{m,d,0} \coord{v}_{m,d,0} + \ldots + \hat{s}_{m,d,R_{m,d}-1} \coord{v}_{m,d,R_{m,d}-1} \\
&=&
\Vm_{m,d} \begin{pmatrix} \hat{s}_{m,d,0} & \ldots & \hat{s}_{m,d,R_{m,d}-1} \end{pmatrix}^T,
\end{eqnarray*}
where $\Vm_{m,d}$ is the block of generalized eigenvectors~\eqref{eq:generalized_eigenvectors_block}. As follows from the Jordan decomposition~\eqref{eq:Jordan_decomposition}, shifting the signal $\coord{s}_{m,d}$ produces the output $\widehat{\coord{s}}_{m,d}\in\S_{m,d}$ from the same subspace, since
\begin{eqnarray}
\nonumber
\widehat{\coord{s}}_{m,d} &=&
\Adj \coord{s}_{m,d} = \Adj \Vm_{m,d} \begin{pmatrix} \hat{s}_{m,d,0} & \ldots & \hat{s}_{m,d,R_{m,d}-1}  \end{pmatrix}^T \\
\nonumber
&=&
\Vm_{m,d} \,\,\Jm_{R_{m,d}}(\lambda_m) \,\begin{pmatrix} \hat{s}_{m,d,0} & \ldots & \hat{s}_{m,d,R_{m,d}-1}  \end{pmatrix}^T \\
\label{eq:shifting_subspace}
&=&
\Vm_{m,d} \begin{pmatrix} \lambda_m \hat{s}_{m,d,0} + \hat{s}_{m,d,1} \\ \vdots \\ \lambda_m \hat{s}_{m,d,R_{m,d}-2} + \hat{s}_{m,d,R_{m,d}-1}  \\ \lambda_m \hat{s}_{m,d,R_{m,d}-1}  \end{pmatrix}.
\end{eqnarray}
Hence, each subspace $\S_{m,d}\leq \S$ is invariant to shifting.

Using~\eqref{eq:Jordan_decomposition} and Theorem~\ref{thm:shift_invariant_polynomials}, we write the graph filter~\eqref{eq:filter} as
\begin{align}
\nonumber
h(\Adj) {}&= \sum_{\ell=0}^{L} h_\ell (\Vm \Jm \Vm^{-1})^\ell
\,\,\,= \sum_{\ell=0}^{L} h_\ell \Vm \Jm^\ell \Vm^{-1} \\
\label{eq:decomposition_h}
{}&= \Vm \big( \sum_{\ell=0}^{L} h_\ell  \Jm^\ell \big) \Vm^{-1}
= \Vm h(\Jm) \Vm^{-1}.
\end{align}
Similarly to~\eqref{eq:shifting_subspace}, we observe that filtering a signal $\coord{s}_{m,d}\in\S_{m,d}$ produces an output $\widehat{\coord{s}}_{m,d}\in\S_{m,d}$ from the same subspace:
\begin{eqnarray}
\nonumber
\widehat{\coord{s}}_{m,d} &=&
h(\Adj) \coord{s}_{m,d} = h(\Adj) \Vm_{m,d} \!\begin{pmatrix} \hat{s}_{m,d,0} \\ \vdots \\ \hat{s}_{m,d,R_{m,d}-1}  \end{pmatrix} \\
\label{eq:filtering_subspace}
&=&
\Vm_{m,d} \,\,\,\left[ \,h(\Jm_{R_{m,d}}(\lambda_m))\,\, \begin{pmatrix} \hat{s}_{m,d,0} \\ \vdots \\ \hat{s}_{m,d,R_{m,d}-1}  \end{pmatrix} \right].
\end{eqnarray}

Since all $N$ generalized eigenvectors of $\Adj$ are linearly independent, all subspaces $\S_{m,d}$ have zero intersections, and their dimensions add to $N$. Thus, the \emph{spectral decomposition}~\eqref{eq:direct_sum} of the signal space $\S$ is
\vspace{-2mm}
\begin{equation}
\label{eq:spectral_decomposition}
\S = \bigoplus_{m=0}^{M-1} \bigoplus_{d=0}^{D_m-1} \S_{m,d}.
\vspace{-2mm}
\end{equation}

\subsection*{Graph Fourier Transform}

The spectral decomposition~\eqref{eq:spectral_decomposition} expands each signal $\coord{s}\in\S$ on the basis of the invariant subspaces of $\S$. Since we chose the generalized eigenvectors as bases of the subspaces $\S_{m,d}$, the expansion coefficients are given by
\begin{equation}
\label{eq:inverse_FT}
\coord{s} = \Vm \coord{\hat{s}},
\end{equation}
where $\Vm$ is the generalized eigenvector matrix~\eqref{eq:generalized_eigenvectors_matrix}.
The vector of expansion coefficients is given by
\begin{equation}
\label{eq:graph_FT}
\coord{\widehat{s}} = \Vm^{-1} \coord{s}.
\end{equation}
The union of the bases of all spectral components~$\S_{m,d}$, i.e., the basis of generalized eigenvectors, is called the \emph{graph Fourier basis}.
We call the expansion~\eqref{eq:graph_FT} of a signal $\coord{s}$ into the graph Fourier basis the \emph{graph Fourier transform} and denote the graph Fourier transform matrix as
\begin{align}\label{eqn:graphFT-1}
\FT = \Vm^{-1}.
\end{align}
Following the conventions of classical DSP, we call the coefficients $\hat{s}_n$ in~\eqref{eq:graph_FT} the \emph{spectrum} of a signal $\coord{s}.$
The \emph{inverse graph Fourier transform} is given by~\eqref{eq:inverse_FT}; it reconstructs the signal from its spectrum.

\subsection*{Frequency Response of Graph Filters}

The \emph{frequency response} of a filter characterizes its effect on the frequency content of the input signal. Let us  rewrite the filtering of $\coord{s}$ by~$h(\Adj)$ using~\eqref{eq:decomposition_h} and~\eqref{eq:inverse_FT} as
\begin{eqnarray}
\nonumber
&& \widetilde{\coord{s}} = h(\Adj)\coord{s} = \FT^{-1} h(\Jm) \FT \coord{s} = \FT^{-1} h(\Jm) \coord{\hat{s}} \\
& \Rightarrow &
\label{eq:conv_theorem}
\FT\widetilde{\coord{s}} = h(\Jm) \coord{\hat{s}}.
\end{eqnarray}
Hence, the spectrum of the output signal is the spectrum of the input signal modified by the block-diagonal matrix
\begin{equation}
\label{eq:frequency_response}
h(\Jm) = \begin{pmatrix}
h(\Jm_{r_{0,0}}(\lambda_0)) \\
& \ddots \\
&& h(\Jm_{r_{M-1,D_M-1}}(\lambda_{M-1}))
\end{pmatrix},
\end{equation}
so that the part of the spectrum corresponding to the invariant subspace $\S_{m,d}$ is multiplied by $h(\Jm_m)$. Hence, $h(\Jm)$ in~\eqref{eq:frequency_response} represents the frequency response of the filter $h(\Adj).$

Notice that~\eqref{eq:conv_theorem} also generalizes the \emph{convolution theorem} from classical DSP~\cite{Oppenheim:99} to arbitrary graphs.
\begin{theorem}
\label{thm:dspg-convolution}
Filtering a signal is equivalent, in the frequency domain, to multiplying its spectrum by the frequency response of the filter.
\end{theorem}

\subsection*{Discussion}

The connection~\eqref{eq:graph_FT} between the graph Fourier transform and the Jordan decomposition~\eqref{eq:Jordan_decomposition} highlights some desirable properties of representation graphs. For graphs with \emph{diagonalizable} adjacency matrices $\Adj$, which have $N$ linearly independent eigenvectors, the frequency response~\eqref{eq:frequency_response} of filters $h(\Adj)$ reduces to a diagonal matrix with the main diagonal containing values $h(\lambda_m)$, where $\lambda_m$ are the eigenvalues of $\Adj$. Moreover, for these graphs, Theorem~\ref{thm:z-transform} provides the closed-form expression~\eqref{eq:condition_on_basis} for the inverse graph Fourier transform $\FT^{-1} = \Vm$. Graphs with symmetric (or Hermitian) matrices, such as undirected graphs, are always diagonalizable and, moreover, have orthogonal graph Fourier transforms: $\FT^{-1}=\FT^H$. This property has significant practical importance, since it yields a closed-form expression~\eqref{eq:condition_on_basis} for $\FT$ and $\FT^{-1}$. Moreover, orthogonal transforms are well-suited for efficient signal representation, as we demonstrate in Section~\ref{sec:Applications}.

\DSPG\ is consistent with the classical DSP theory. As mentioned in Section~\ref{sec:Signals_on_graphs}, finite discrete periodic time series are represented by the directed graph in Fig.~\ref{fig:graphs_time}. The corresponding adjacency matrix is the $N\times N$ circulant matrix~\eqref{eq:time_shift_matrix}.
Its eigendecomposition (and hence, Jordan decomposition) is
\begin{equation*}
\Circ_N =
\frac{1}{N}
\DFT_N^{-1}
\begin{pmatrix}
e^{-j\frac{2\pi \cdot 0}{N}} \\
& \ddots \\
&& e^{-j\frac{2\pi\cdot (N-1)}{N}}
\end{pmatrix}
\DFT_N,
\end{equation*}
where $\DFT_N$ is the discrete Fourier transform matrix. Thus, as expected, the graph Fourier transform is $\FT=\DFT_N$. Furthermore, for a general filter $h(\Circ_N) = \sum_{\ell=0}^{N-1} h_\ell \Circ_N^\ell$, coefficients of the output $\widehat{\coord{s}} = h(\Circ_N) \coord{s}$ are calculated as
\begin{eqnarray*}
\widehat{s}_n &=& h_n s_0 + \ldots + h_0 s_n + h_{N-1} s_{n+1} + \ldots + h_{n+1} s_{N-1} \\
&=& \sum_{k=0}^{N-1} s_k h_{(n-k \mod N)}.
\end{eqnarray*}
This is the standard circular convolution. Theorem~\ref{thm:impulseresponse-1} holds as well, with impulse response identical to filter taps: $\coord{u} = \coord{h}$.

Similarly, it has been shown in~\cite{Pueschel:08b,Pueschel:05e} that unweighted line graphs similar to Fig.~\ref{fig:graphs_time}, but with undirected edges
and different, non-periodic boundary conditions, give rise to all $16$ types of discrete cosine and sine transforms as their graph Fourier transform matrices.
Combined with~\cite{Dudgeon:83}, it can be shown that graph Fourier transforms for images on the lattice in Fig.~\ref{fig:graphs_image} are different types of two-dimensional discrete cosine and sine transforms, depending on boundary conditions. This result serves as additional motivation for the use of these transforms in image representation and coding~\cite{Bovik:05}.

In discrete-time DSP, the concepts of filtering, spectrum, and Fourier transform have natural, physical interpretations. In \DSPG, when instantiated for various datasets, the interpretation of these concepts may be drastically different and not immediately obvious. For example, if a representation graph reflects the proximity of sensors in some metric (such as time, space, or geodesic distance), and the dataset contains sensor measurements, then filtering corresponds to linear recombination of related measurements and can be viewed as a graph form of regression analysis with constant coefficients. The graph Fourier transform then decomposes signals over equilibrium points of this regression. On the other hand, if a graph represents a social network of individuals and their communication patterns, and the signal is a social characteristic, such as an opinion or a preference, then filtering can be viewed as diffusion of information along established communication channels, and the graph Fourier transform characterizes signals in terms of stable, unchangeable opinions or preferences.

\section{Applications}
\label{sec:Applications}

We consider several applications of \DSPG\ to data processing. These examples illustrate the effectiveness of the framework in standard DSP tasks,
such as predictive filtering and efficient data representation, as well as demonstrate that the framework can tackle problems less common in DSP,
such as data classification and customer behavior prediction.

\subsection*{Linear Prediction}

\begin{figure}
  \begin{center}
    \subfigure[Forward (prediction) filter]{\includegraphics[scale=0.5]{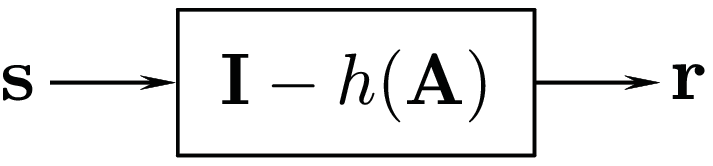}}
    \quad \quad
    \subfigure[Backward (synthesis) filter]{\includegraphics[scale=0.5]{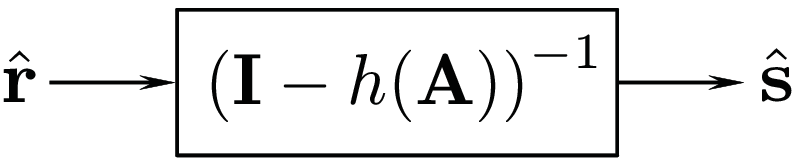}}
  \end{center}
\caption{\label{fig:LPC}
Components of linear prediction.}
\end{figure}

Linear prediction (LP) is an efficient technique for representation, transmission, and generation of time series~\cite{Vaidyanathan:08}. It is used in many applications, including power spectral estimation and direction of arrival analysis. Two main steps of LP are the construction of a prediction filter and the generation of an (approximated) signal, implemented, respectively, with forward and backward filters, shown in Fig.~\ref{fig:LPC}. The forward (prediction) filter converts the signal into a \emph{residual}, which is then closely approximated, for example, by a white noise--flat power spectrum signal or efficient quantization with few bits. The backward (synthesis) filter constructs an approximation of the original signal from the approximated residual.

Using the \DSPG, we can extend LP to graph signals. We illustrate it with the dataset~\cite{NCDC} of daily temperature measurements from sensors located near 150 major US cities. Data from each sensor is a separate time series, but encoding it requires buffering measurements from multiple days before they can be encoded for storage or transmission. Instead, we build a LP filter on a graph to encode daily snapshots of all 150 sensor measurements.

We construct a representation graph $G=(\Nodes,\Adj)$ for the sensor measurements using geographical distances between sensors. Each sensor corresponds to a node $v_n$, $0\leq n < 150$, and is connected to $K$ nearest sensors with undirected edges weighted by the normalized inverse exponents of the squared distances: if $d_{nm}$ denotes the distance between the $n$th and $m$th sensors\footnote{The construction of representation graphs for datasets is an important research problem and deserves a separate discussion that is beyond the scope of this paper. The procedure we use here is a popular choice for construction of similarity graphs based on distances between nodes~\cite{Belkin:03,Coifman:06,Hammond:11}.} and $m\in\Neighb_n$, then
\begin{equation}
\label{eq:A_temperature}
A_{n,m} = \frac{e^{-d_{nm}^2}}{\sqrt{\sum_{k\in\Neighb_n}e^{-d_{nk}^2} \sum_{\ell\in\Neighb_m}e^{-d_{m\ell}^2}}}.
\end{equation}

\begin{figure}
  \begin{center}
      \vspace{-20mm}
      \includegraphics[scale=0.4]{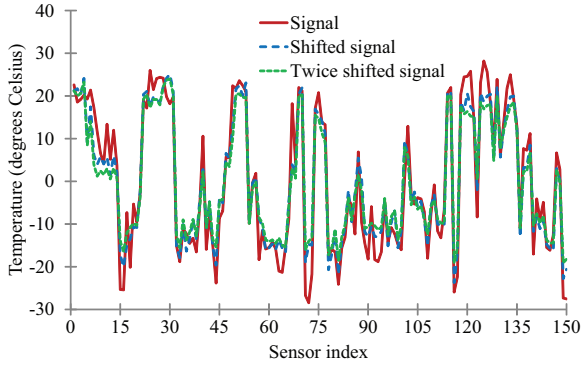}
  \end{center}
\caption{\label{fig:shifted_signal}
A signal representing a snapshot of temperature measurements from $N=150$ sensors.
Shifting the signal produces signals similar to the original.}
\end{figure}

For each snapshot $\coord{s}$ of $N=150$ measurements, we construct a prediction filter $h(\Adj)$ with $L$ taps by minimizing the energy of the residual
$\coord{r} = \coord{s} - h(\Adj)\coord{s} = \left(\Id_N - h(\Adj)\right)\coord{s}$. We set $h_0=0$ to avoid the trivial solution $h(\Adj)=\Id$, and obtain
\begin{equation*}
\begin{pmatrix} h_1 & \ldots & h_{L-1} \end{pmatrix}^T
= (\coord{B}^T\coord{B})^{-1}\coord{B}^T\coord{s}.
\end{equation*}
Here,
$\coord{B} = \begin{pmatrix}
\Adj\coord{s} & \ldots & \Adj^{L-1}\coord{s}
\end{pmatrix}$ is a $N\times (L-1)$ matrix.
The residual energy $||\coord{r}||_2^2$ is relatively small compared to the energy of the signal $\coord{s}$, since shifted signals are close
approximations of $\coord{s}$, as illustrated in Fig.~\ref{fig:shifted_signal}. This phenomenon provides the intuition for the graph shift: if the graph represents a similarity relation, as in this example, then the shift replaces each signal sample with a sum of related samples with more similar samples weighted heavier than less similar ones.


\begin{figure}
  \begin{center}
      \includegraphics[scale=0.45]{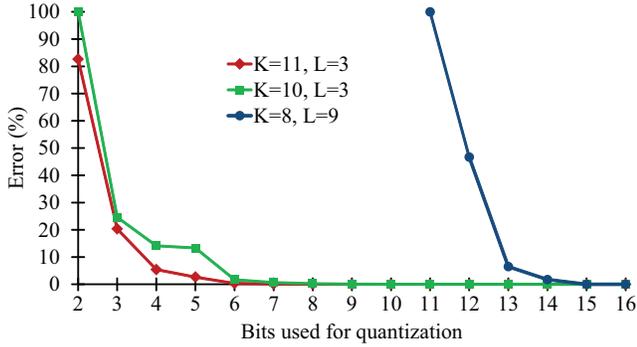}
  \end{center}
\caption{\label{fig:lpc_error} Average approximation errors $||\coord{s}-\widehat{\coord{s}}||_2/||\coord{s}||_2$ for LP coding of $365$ signals $\coord{s}$ representing daily temperature snapshots. Graphs with $1\leq K \leq 15$ nearest neighbors for each sensor were analyzed, and filters with $2\leq L \leq 10$ taps were constructed. The residual was quantized using $1\leq B \leq 16$ bits. The lowest, second lowest, and highest errors were obtained, respectively for $K=11$ and $L=3$,  $K=10$ and $L=3$, and $K=8$ and $L=9$.}
\end{figure}

The residual $\coord{r}$ is then quantized using $B$ bits, and the quantized residual $\widehat{\coord{r}}$ is processed with the inverse filter to
synthesize an approximated signal $\widehat{\coord{s}}=\left(\Id_N - h(\Adj)\right)^{-1} \widehat{\coord{r}}$.

We considered graphs with $1\leq K \leq 15$ nearest neighbors, and for each~$K$ constructed optimal prediction filters with $2\leq L \leq 10$ taps.
As shown in Fig.~\ref{fig:lpc_error}, the lowest and highest errors $||\coord{s}-\widehat{\coord{s}}||_2/||\coord{s}||_2$ were obtained for $K=11$ and $L=3$,
and for $K=8$ and $L=9$. During the experiments, we observed that graphs with few neighbors (approximately, $3\leq K\leq 7$) lead to lower
errors when prediction filters have impulse responses of medium length ($4\leq L \leq 6$), while graphs with $7\leq K\leq 11$ neighbors yield lower
errors for $3\leq L\leq 5$. Using larger values of~$K$ and~$L$ leads to large errors. This tendency may be due to overfitting filters to signals, and
demonstrates that there exists a trade-off between graph and filter parameters.

\subsection*{Signal Compression}

Efficient signal representation is required in multiple DSP areas, such as storage, compression, and transmission. Some widely-used techniques are based on expanding signals into orthonormal bases with the expectation that most information is captured with few basis functions. The expansion coefficients are calculated using an orthogonal transform. If the transform represents a Fourier transform in some model, it means that signals are sparse in the frequency domain in this model, i.e., they contain only few frequencies. Some widely-used image compression standards, such as JPEG and JPEG~2000, use orthogonal expansions implemented, respectively, by discrete cosine and wavelet transforms~\cite{Bovik:05}.

As discussed in the previous example, given a signal $\coord{s}$ on a graph $G=(\Nodes,\Adj)$, where $\Adj$ reflects similarities between data elements, the shifted signal $\Adj \coord{s}$ can be a close approximation of $\coord{s}$, up to a scalar factor: $\Adj \coord{s} \approx \rho \coord{s}$. This is illustrated in Fig.~\ref{fig:shifted_signal}, where $\rho\approx 1$. Hence, $\coord{s}$ can be effectively expressed as a linear combination of a few [generalized] eigenvectors of~$\Adj$.

\begin{figure}
  \begin{center}
      \includegraphics[scale=0.4]{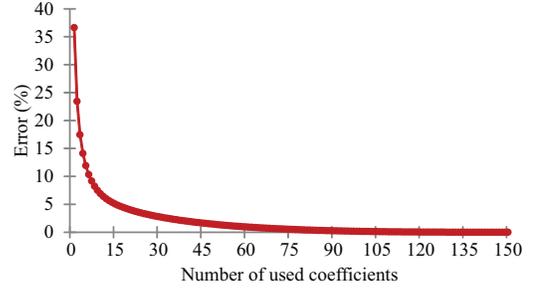}
  \end{center}
\caption{\label{fig:FT_error}
Average reconstruction error $||\coord{s}-\coord{\tilde{s}}||_2/||\coord{s}||_2$ for the compression of $365$ daily temperature snapshots based on the graph Fourier transform using $1\leq C \leq N$ coefficients.}
\end{figure}

Consider the above dataset of temperature measurements. The matrix $\Adj$ in~\eqref{eq:A_temperature} is symmetric by construction, hence its eigenvectors form an orthonormal basis, and the graph Fourier transform matrix $\FT$ is orthogonal. In this case, we can compress each daily update $\coord{s}$ of $N=150$ measurements by keeping only the $C$ spectrum coefficients~\eqref{eq:graph_FT} $\coord{\hat{s}}_n$ with largest magnitudes. Assuming that $|\coord{\hat{s}}_0| \geq |\coord{\hat{s}}_1| \geq \ldots \geq |\coord{\hat{s}}_{N-1}|$, the signal reconstructed after compression is
\begin{equation}
\label{eq:S_compressed}
\coord{\tilde{s}} = \FT^{-1} \left( \coord{\hat{s}}_0, \ldots , \coord{\hat{s}}_{C-1}, 0 , \ldots , 0\right)^T.
\end{equation}
Fig.~\ref{fig:FT_error} shows the average reconstruction errors obtained by retaining $1\leq C \leq N$ spectrum coefficients.

\begin{figure}
  \begin{center}
    \includegraphics[scale=0.25]{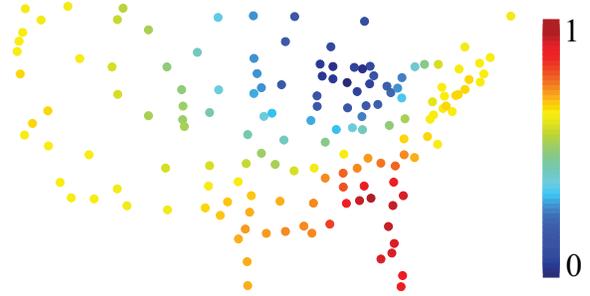}
  \end{center}
\caption{\label{fig:weather_eigenvector}
The Fourier basis vector that captures most energy of temperature measurements
reflects the relative distribution of temperature across the mainland United States.
The coefficients are normalized to the interval $[0,1]$.}
\end{figure}

This example also provides interesting insights into the temperature distribution pattern in the United States. Consider the Fourier basis vector that most frequently (for 217 days out of 365) captures most energy of the snapshot $\coord{s}$, i.e., yields the spectrum coefficient $\coord{\hat{s}}_0$ in~\eqref{eq:S_compressed}. Fig.~\ref{fig:weather_eigenvector} shows the vector coefficients plotted on the representation graph according to the sensors' geographical coordinates, so the graph naturally takes the shape of the mainland US. It can be observed that this basis vector reflects the relative temperature distribution across the US: the south-eastern region is the hottest one, and the Great Lakes region is the coldest one~\cite{NCDC11}.

\subsection*{Data Classification}

Classification and labeling are important problems in data analysis. These problems have traditionally been studied in machine learning~\cite{Mitchell:97,Duda:00}. Here, we propose a novel data classification algorithm by demonstrating that a classifier system can be interpreted as a filter on a graph. Thus, the construction of an optimal classifier can be viewed and studied as the design of an adaptive graph filter. Our algorithm scales linearly with the data size $N$, which makes it an attractive alternative to existing classification methods based on neural networks and support vector machines.

Our approach is based on the label propagation~\cite{Zhu:03,Wang:06}, which is a simple, yet efficient technique that advances known labels from labeled graph nodes along edges to unlabeled nodes. Usually this propagation is modeled as a stationary discrete-time Markov process~\cite{Papoulis:02}, and the graph adjacency matrix is constructed as a probability transition matrix, i.e., $\Adj_{n,m} \geq 0$ for all $n,m$, and $\Adj \OneVector_N = \OneVector_N$, where $\OneVector_N$ is a column vector of $N$ ones. Initially known labels determine the initial probability distribution $\coord{s}$. For a binary classification problem with only two labels, the resulting labels are determined by the distribution $\coord{\tilde{s}} = \Adj^P \coord{s}.$ If $\tilde{s}_n \leq 1/2$, node $v_n$ is assigned one label, and otherwise the other. The number $P$ of propagations is determined heuristically.

Our \DSPG\ approach has two major distinctions from the original label propagation. First, we do not require $\Adj$ to be a stochastic matrix. We only assume that edge weights $\Adj_{k,m}\geq 0$ are non-negative and indicate similarity or dependency between nodes. In this case, nodes with positive labels $\tilde{s}_n>0$ are assigned to one class,
and with negative labels to another. Second, instead of propagating labels as in a Markov chain, we construct a filter $h(\Adj)$ that produces labels
\begin{equation}
\label{eq:labels}
\coord{\tilde{s}} = h(\Adj) \coord{s}.
\end{equation}

The following example illustrates our approach. Consider a set of $N=1224$ political blogs on the Web that we wish to classify as ``conservative''
or ``liberal'' based on their context~\cite{Adamic:05}. Reading and labeling each blog is very time-consuming. Instead, we read and label only a few blogs,
and use these labels to adaptively build a filter $h(\Adj)$ in~\eqref{eq:labels}.

Let signal $\coord{s}$ contain initially known labels, where ``conservative,'' ``liberal,'' and unclassified blogs are represented by values $s_n=+1$, $-1$, and $0$, respectively. Also, let signal $\coord{t}$ contain \emph{training} labels, a subset of known labels from $\coord{s}$. Both $\coord{s}$ and $\coord{t}$ are represented by a graph $G=(\Nodes, \Adj)$, where node $v_n$ containing the label of the $n$th blog, and edge $\Adj_{n,m}=1$ if and only if there is a hyperlink reference from the $n$th to the $m$th blog; hence the graph is directed. Observe that the discovery of hyperlink references is a fast, easily automated task, unlike reading the blogs. An example subgraph for $50$ blogs is shown in Fig.~\ref{fig:graphs_blogs}.

Recall that the graph shift $\Adj$ replaces each signal coefficient with a weighted combination of its neighbors. In this case, processing training labels $\coord{t}$ with the filter
\begin{equation}
\label{eq:h_step}
\Id_N + h_1 \Adj
\end{equation}
produces new labels $\coord{\tilde{t}}=\coord{t} + h_1 \Adj\coord{t}.$ Here, every node label is adjusted by a scaled sum of its neighbors' labels.
The parameter $h_1$ can be interpreted as the ``confidence'' in our knowledge of current labels: the higher the confidence $h_1$, the more neighbors' labels
should affect the current labels. We restrict the value of $h_1$ to be positive.

Since the sign of each label indicates its class, label $\tilde{t}_n$ is incorrect if its sign differs from $s_n$, or $\tilde{t}_n s_n \leq 0$ for $s_n\neq 0$.
We determine the optimal value of $h_1$ by minimizing the total error, given by the number of incorrect and undecided labels.
This is done in linear time proportional to the number of initially known labels $s_n\neq 0$, since each constraint
\begin{equation}
\label{eq:label_constraint}
\tilde{t}_n s_n  = \left( t_n + h_1 \sum_{k\in\Neighb_n} t_k\right)s_n \leq 0
\end{equation}
is a linear inequality constraint on $h_1$.

To propagate labels to all nodes, we repeatedly feed them through $P$ filters~\eqref{eq:h_step} of the form $h^{(p)}(\Adj) = \Id_N + h_p \Adj$, each time optimizing
the value of $h_p$ using the greedy approach discussed above. The obtained adaptive classification filter is
\begin{equation}
\label{eq:classifier_filter}
h(\Adj) = (\Id_N + h_P \Adj)(\Id_N + h_{P-1} \Adj)\ldots(\Id_N + h_1 \Adj).
\end{equation}
In experiments, we set $P=10$, since we observed that filter~\eqref{eq:classifier_filter} converges
quickly, and in many cases, $h_p=0$ for $p>10$, which is similar to the actual graph's diameter of $8$.
After the filter~\eqref{eq:classifier_filter} is constructed, we apply it to all known labels $\coord{s},$
and classify all $N$ nodes based on the signs of resulting labels $\coord{\tilde{s}} = h(\Adj)\coord{s}$.

In our experiments, we considered two methods for selecting nodes to be labeled initially: random selection, and selection of blogs with most hyperlinks.
As Table~\ref{tab:classification} shows, our algorithm achieves high accuracy for both methods.
In particular, assigning initial labels $\coord{s}$ to only $2\%$ of blogs with most hyperlinks leads to the correct classification of $93~\%$ of unlabeled blogs.
%

\begin{table}
\centering
\footnotesize
\ra{1.5}
\begin{tabular}{@{}llll@{}}
\toprule
\multirow{2}{*}{Blog selection method}& \multicolumn{3}{c}{Fraction of initially known labels} \\
& $2\%$ & $5\%$ & $10\%$ \\
\midrule
Random & $87\%$ & $93\%$ & $95\%$ \\
Most hyperlinks & $93\%$ & $94\%$ & $95\%$ \\
\bottomrule
\end{tabular}
\caption{\label{tab:classification}
Accuracy of blog classification using adaptive filters.}
\end{table}

\subsection*{Customer Behavior Prediction}
The adaptive filter design discussed in the previous example can be applied to other problems as well. Moreover, the linear computational cost of the filter design makes the approach easily scalable for the analysis of large datasets. Consider an example of a mobile service provider that is interested in keeping its customers. The company wants to predict which users will stop using their services in the near future, and offer them incentives for staying with the provider (improved call plan, discounted phones, etc.). In particular, based on their past behavior, such as number and length of calls within the network, the company wants to predict whether customers will stop using the services in the next month.

This problem can be formulated similarly to the previous example. In this case, the value at node $v_n$ of the representation graph $G=(\Nodes,\Adj)$ indicates the probability that the $n$th customer will not use the provider services in the next 30 days. The weight of a directed edge from node $v_n$ to $v_m$ is the fraction of time the $n$th customer called and talked to the $m$th customer; i.e., if $T_{n,m}$ indicates the total time the $n$th customer called and talked to the $m$th customer in the past until the present moment, then
$$
\Adj_{n,m} = \frac{T_{n,m}}{\sum_{k\in\Neighb_n}T_{n,k}}.
$$
The initial input signal $\coord{s}$ has $s_n=1$ if the customer has already stopped using the provider, and $s_n=0$ otherwise. As in the previous example, we design a classifier filter~\eqref{eq:classifier_filter}; we set $P=10$. We then process the entire signal $\coord{s}$ with the designed filter obtaining the output signal $\coord{\tilde{s}}$ of the predicted probabilities; we conclude that the $n$th customer will stop using the provider if $\tilde{s}_n\geq 1/2$, and will continue if $\tilde{s}_n<1/2$.

\begin{figure}
  \begin{center}
    \includegraphics[scale=0.4]{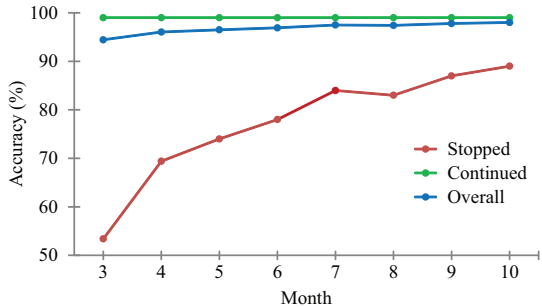}
  \end{center}
\vspace*{-3mm}
\caption{\label{fig:mobile}
The accuracy of behavior prediction for customers of a mobile provider. Predictions for customers who stopped using the provider and those who continued are evaluated separately, and then combined into the overall accuracy.}
\end{figure}

In our preliminary experiments, we used a ten-month-long call log for approximately $3.5$ million customers of a European mobile service provider, approximately $10\%$ of whom stopped using the provider during this period\footnote{We use a large dataset on Call Detailed Records (CDRs) from a large mobile operator in one European country, which we call EURMO for short.}. Fig.~\ref{fig:mobile} shows the accuracy of predicting customer behavior for months 3-10 using filters with at most $L \leq 10$ taps. The accuracy reflects the ratio of correct predictions for all customers, the ones who stop using the service and the ones who continue; it is important to correctly identify both classes, so the provider can focus on the proper set of customers. As can be seen from the results, the designed filters achieve high accuracy in the prediction of customer behavior. Unsurprisingly, the prediction accuracy increases as more information becomes available, since we optimize the filter for month $K$ using cumulative information from preceding $K-1$ months.

\section{Conclusions}

We have proposed \DSPG, a novel DSP theory for datasets whose underlying similarity or dependency relations are represented by arbitrary graphs.
Our framework extends fundamental DSP structures and concepts, including shift, filters, signal and filter spaces,
spectral decomposition, spectrum, Fourier transform, and frequency response, to such datasets by viewing them as signals indexed by graph nodes.
We demonstrated that \DSPG\ is a natural extension of the classical time-series DSP theory, and traditional definitions of the above DSP concepts
and structures can be obtained using a graph representing discrete time series.
We also provided example applications of \DSPG\ to various social science applications, and our experimental results demonstrated the effectiveness
of using the \DSPG\ framework for datasets of different nature.

\subsection*{Acknowledgment}
We thank EURMO, CMU Prof.~Pedro Ferreira, and the iLab at CMU Heinz College for granting us access to EURMO CDR database and related discussions.
%

\section*{Appendix~A: Matrix Decomposition and Properties}
\label{app:jordanminimalpoly}
We review relevant properties of the Jordan normal form,
and the characteristic and minimal polynomial of a matrix $\Adj\in\C^{N\times N}$; for a thorough review, see~\cite{Gantmacher:59,Lancaster:85}.

\subsection*{Jordan Normal Form}
\label{app:jordannormalform}
Let $\lambda_0,\ldots,\lambda_{M-1}$ denote $M\leq N$ distinct eigenvalues of $\Adj$. Let each eigenvalue $\lambda_m$ have $D_m$ linearly independent eigenvectors $\coord{v}_{m,0}, \ldots,\coord{v}_{m,D_m-1}$. The $D_m$ is the \textit{geometric multiplicity} of~$\lambda_m$. Each eigenvector $\coord{v}_{m,d}$ generates a \emph{Jordan chain} of $R_{m,d}\geq 1$ linearly independent \emph{generalized eigenvectors} $\coord{v}_{m,d,r}$, $0\leq r < R_{m,d}$, where $\coord{v}_{m,d,0}=\coord{v}_{m,d}$, that satisfy
\begin{equation}
\label{eq:generalized_eigenvector_condition}
(\Adj-\lambda_m\Id)\coord{v}_{m,d,r}=\coord{v}_{m,d,r-1}.
\end{equation}

For each eigenvector $\coord{v}_{m,d}$ and its Jordan chain of length $R_{m,d}$, we define a \emph{Jordan block} matrix of dimension~$R_{m,d}$ as
\begin{equation}
\label{eq:Jordan_block}
J_{r_{m,d}}(\lambda_m) = \begin{pmatrix}
\lambda_m & 1 \\
& \lambda_m & \ddots \\
&& \ddots & 1 \\
&&& \lambda_m
\end{pmatrix} \in \C^{R_{m,d}\times R_{m,d}}.
\end{equation}
Thus, each eigenvalue $\lambda_m$ is associated with~$D_m$ Jordan blocks, each with dimension $R_{m,d}$, $0\leq d < D_m.$
Next, for each eigenvector $\coord{v}_{m,d}$, we collect its Jordan chain into a $N\times R_{m,d}$ matrix
\begin{equation}
\label{eq:generalized_eigenvectors_block}
\Vm_{m,d} = \begin{pmatrix}
\coord{v}_{m,d,0} & \ldots & \coord{v}_{m,d,R_{m,d}-1}
\end{pmatrix}.
\end{equation}
We concatenate all blocks $\Vm_{m,d}$, $0\leq d< D_m$ and $0\leq m<M$, into one block matrix
\begin{equation}
\label{eq:generalized_eigenvectors_matrix}
\Vm = \begin{pmatrix}
\Vm_{0,0} & \ldots & \Vm_{M-1,D_{M-1}}
\end{pmatrix},
\end{equation}
so that $\Vm_{m,d}$ is at position $\sum_{k=0}^{m-1}D_k+d$ in this matrix.
Then, matrix $\Adj$ has the \emph{Jordan decomposition}
\begin{equation}
\label{eq:Jordan_decomposition}
\Adj = \Vm \Jm \Vm^{-1},
\end{equation}
where the block-diagonal matrix
\begin{equation}
\label{eq:Jordan_normal_form}
\Jm = \begin{pmatrix}
\Jm_{R_{0,0}}(\lambda_0) \\
& \ddots \\
&& \Jm_{R_{M-1,D_{M-1}}}(\lambda_{M-1})
\end{pmatrix}
\end{equation}
is called the \emph{Jordan normal form} of $\Adj$.

\subsection*{Minimal and Characteristic Polynomials}
\label{subsec:minpoly}
The \emph{minimal polynomial} of matrix $\Adj$ is the monic polynomial of smallest possible degree that satisfies $m_\Adj(\Adj) = \ZeroMatrix_N$. Let $R_m = \max\{ R_{m,0}, \ldots, R_{m,D_m-1}\}$ denote the maximum length of Jordan chains corresponding to eigenvalue $\lambda_m$. Then the minimal polynomial $m_\Adj(x)$ is given by
\begin{equation}
\label{eq:minimal_polynomial}
m_\Adj(x) = (x-\lambda_0)^{R_1} \ldots (x-\lambda_{M-1})^{R_{M-1}}.
\end{equation}
The \textit{index} of~$\lambda_m$ is $R_m$, $1\leq m<M$.
Any polynomial $p(x)$ that satisfies $p(\Adj)=\ZeroMatrix_N$, is a polynomial multiple of $m_\Adj(x)$, i.e., $p(x) = q(x)m_\Adj(x)$.
The degree of the minimal polynomial satisfies
\begin{equation}
\label{eq:degree_minimal_polynomial}
\deg m_\Adj(x) = N_{\Adj}=\sum_{m=0}^{M-1}\,R_m \leq N.
\end{equation}

The \emph{characteristic polynomial} of the matrix $\Adj$ is defined as
\begin{equation}
\label{eq:characteristic_polynomial}
p_\Adj(x) = \det(\lambda\Id - \Adj) = (x-\lambda_0)^{A_0} \ldots (x-\lambda_{M-1})^{A_{M-1}}.
\end{equation}
Here: $A_m = R_{m,0} + \ldots + R_{m,D_m-1}$ for $0\leq m < M$, is the \textit{algebraic multiplicity} of~$\lambda_m$; $\deg p_\Adj(x)=N$; $p_\Adj(x)$ is a multiple of $m_\Adj(x)$; and $p_\Adj(x)=m_\Adj(x)$ if and only if the geometric multiplicity of each~$\lambda_m$, $D_m=1$, i.e., each eigenvalue $\lambda_m$ has exactly one eigenvector.

\section*{Appendix B: Proof of Theorem~\ref{thm:shift_via_shift}}
We will use the following lemma to prove Theorem~\ref{thm:shift_via_shift}.
\begin{lemma}
\label{lemma:hgJ}
For polynomials $h(x)$, $g(x)$, and $p(x)=h(x)g(x)$, and a Jordan block~$J_r(\lambda)$ as in~\eqref{eq:Jordan_block} of arbitrary dimension~$r$ and eigenvalue~$\lambda$, the following equality holds:
\begin{equation}
\label{eq:hgJ_equals_pJ}
h(\Jm_r(\lambda)) g(\Jm_r(\lambda)) = p(\Jm_r(\lambda)).
\end{equation}
\end{lemma}
\begin{IEEEproof}
The $(i,j)$th element of $h(\Jm_r(\lambda))$ is
\begin{equation}
\label{eq:hJ_ij}
h(\Jm_r(\lambda))_{i,j} = \frac{1}{(j-i)!}h^{(j-i)}(\lambda)
\end{equation}
for $j\geq i$ and $0$ otherwise, where $h^{(j-i)}(\lambda)$ is the $(j-i)$th derivative of $h(\lambda)$~\cite{Lancaster:85}.
Hence, the $(i,j)$th element of $h(\Jm_r(\lambda)) g(\Jm_r(\lambda))$
for $j<i$ is zero and for $j\geq i$ is
\begin{eqnarray}
\nonumber
&&
\sum_{k=i}^{j} h(\Jm_r(\lambda))_{i,k} g(\Jm_r(\lambda))_{k,j} \\
\nonumber
&=&
\sum_{k=i}^{j} \frac{1}{(k-i)!}h^{(k-i)}(\lambda) \frac{1}{(j-k)!}g^{(j-k)}(\lambda) \\
\nonumber
&=&
\frac{1}{(j-i)!}
\sum_{k=i}^{j} { j-i \choose k-i } h^{(k-i)}(\lambda) g^{(j-k)}(\lambda) \\
\nonumber
&=&
\frac{1}{(j-i)!}
\sum_{m=0}^{j-i} { j-i \choose m } h^{(m)}(\lambda) g^{(j-i-m)}(\lambda) \\
\label{eq:hgJ_ij}
&=&
\frac{1}{(j-i)!} \big(h(\lambda) g(\lambda)\big)^{(j-i)}.
\end{eqnarray}
Matrix equality~\eqref{eq:hgJ_equals_pJ} follows by comparing~\eqref{eq:hgJ_ij} with~\eqref{eq:hJ_ij}.
\end{IEEEproof}

As before, let $\lambda_0,\ldots, \lambda_{M-1}$ denote distinct eigenvalues of $\Adj$. Consider the Jordan decomposition~\eqref{eq:Jordan_decomposition} of $\Adj$.
For each $0\leq m<M$, select distinct numbers $\widetilde{\lambda}_{m,0},\ldots,\widetilde{\lambda}_{m,D_m-1}$, so that all $\widetilde{\lambda}_{m,d}$ for $0\leq d< D_m$ and $0\leq m < M$ are distinct. Construct the block-diagonal matrix
\begin{equation*}
\widetilde{\Jm} = \begin{pmatrix}
\Jm_{R_{0,0}}(\widetilde{\lambda}_{0,0}) \\
& \ddots \\
&& \Jm_{R_{M-1,D_{M-1}}}(\widetilde{\lambda}_{M-1,D_{M-1}-1})
\end{pmatrix}.
\end{equation*}
The Jordan blocks on the diagonal of $\widetilde{\Jm}$ match the sizes of the Jordan blocks of $\Jm$ in~\eqref{eq:Jordan_normal_form}, but their elements are different.

Consider a polynomial $r(x) = r_0 + r_1 x + \ldots + r_{N-1}x^{N-1}$, and assume that $r(\widetilde{\Jm}) = \Jm$. By Lemma~\ref{lemma:hgJ}, this is equivalent to
\begin{equation*}
\begin{cases}
r(\widetilde{\lambda}_{m,d}) = \lambda_m ,\\
r^{(1)}(\widetilde{\lambda}_{m,d}) = 1 \\
r^{(i)}(\widetilde{\lambda}_{m,d}) = 0,\text{   for   } 2\leq i < D_m
\end{cases}
\end{equation*}
for all $0\leq d< D_m$ and $0\leq m < M$. This is a system of $N$ linear equations with $N$ unknowns $r_0,\ldots,r_{N-1}$ that can be uniquely solved using inverse polynomial interpolation~\cite{Lancaster:85}.

Using~\eqref{eq:Jordan_decomposition}, we obtain $\Adj = \Vm\Jm\Vm^{-1} = \Vm r(\widetilde{\Jm})\Vm^{-1} = r(\Vm \widetilde{\Jm} \Vm^{-1}) = r(\widetilde{\Adj}).$
Furthermore, since all $\widetilde{\lambda}_{m,d}$ are distinct numbers, their geometric multiplicities are equal to $1$. As discussed in Appendix~A,
this is equivalent to $p_{\tilde{\Adj}}(x) = m_{\tilde{\Adj}}(x).$

\section*{Appendix C: Proof of Theorem~\ref{thm:inverse_filter}}

Lemma~\ref{lemma:hgJ} leads to the construction procedure of the inverse polynomial $g(x)$ of $h(x)$, when it exists,
and whose matrix representation satisfies 
 $g(\Adj)h(\Adj)=\Id_N$. Observe that this condition, together with~\eqref{eq:hgJ_equals_pJ}, is equivalent to
\begin{equation}
\label{eq:system_for_hg}
\begin{cases}
h(\lambda_m)g(\lambda_m) = 1 ,\text{  for  } 0\leq m\leq M-1\\
\big(h(\lambda_m)g(\lambda_m)\big)^{(i)} = 0,\text{   for   } 1\leq i < R_m.
\end{cases}
\end{equation}
Here, $R_m$ is the degree of the factor $(x-\lambda_m)^{R_m}$ in the minimal polynomial $m_\Adj(\lambda)$ in~\eqref{eq:minimal_polynomial}.
Since values of $h(x)$ and its derivatives at $\lambda_m$ are known, \eqref{eq:system_for_hg} amount to $N_\Adj$ linear equations with $N_\Adj$ unknowns.
They have a unique solution if and only if $h(\lambda_m)\neq 0$ for all $\lambda_m$, and the coefficients $g_0,\ldots,g_{M_{\Adj}-1}$ are
then uniquely determined using inverse polynomial interpolation~\cite{Lancaster:85}.
%
%
\bibliographystyle{IEEEbib}
\begin{small}
\bibliography{references}
\end{small}
\end{document}